\documentclass[9pt,twocolumn,twoside]{pnas-new-anshuman}

\templatetype{pnasresearcharticle} 

\usepackage{multirow}
\usepackage{hyperref}
\makeatletter
\newcommand{\dummylabel}[2]{\def\@currentlabel{#2}\label{#1}}
\makeatother

\title{Incentivizing News Consumption on Social Media Platforms Using Large Language Models and Realistic Bot Accounts}

\author[a]{Hadi Askari}
\author[a,1]{Anshuman Chhabra} 
\author[b]{Bernhard Clemm von Hohenberg}
\author[c, 1]{Michael Heseltine}
\author[a, c, 1, 2]{Magdalena Wojcieszak}

\affil[a]{University of California, Davis}
\affil[b]{GESIS – Leibniz-Institute for the Social Sciences}
\affil[c]{University of Amsterdam}

\authorcontributions{H.A., A.C., B.C., and M.W. designed the research, A.C., B.C., M.H., H.A. developed the pipeline for and oversaw user collection, experimental assignment, behavioral data collection, A.C., B.C., and H.A. developed the tools (A.C. developed the classifiers/LLM response module and H.A. the keywords), H.A. and M.H. developed the bots, M.H. did the analyses, M.W. wrote the paper.}
\authordeclaration{The authors declare no competing interests here.}
\equalauthors{\textsuperscript{1}The authors contributed equally to this work; author order is alphabetical.}
\correspondingauthor{\textsuperscript{2}To whom correspondence should be addressed. E-mail: mwojcieszak@ucdavis.edu}


\begin{abstract}
Polarization, declining trust, and wavering support for democratic norms are pressing threats to U.S. democracy. Exposure to verified and quality news may lower individual susceptibility to these threats and make citizens more resilient to 
misinformation, populism, and hyperpartisan rhetoric. This project examines how to enhance users' exposure to and engagement with verified and ideologically balanced news in an ecologically valid setting. We rely on a large-scale two-week long field experiment (from 1/19/2023 to 2/3/2023) on 28,457 Twitter users. We created 28 bots utilizing GPT-2 that replied to users tweeting about sports, entertainment, or lifestyle with a contextual reply containing two hardcoded elements: a URL to the topic-relevant section of quality news organization and an encouragement to follow its Twitter account. To further test differential effects by gender of the bots, treated users were randomly assigned to receive responses by bots presented as female or male. We examine whether our over-time intervention enhances the following of news media organization, the sharing and the liking of news content (as determined by our extensive curated list of news media organizations) and the tweeting about politics and the liking of political content (as determined using our fine-tuned RoBERTa NLP transformer-based model). We find that the treated users followed more news accounts and the users in the female bot treatment were more likely to like news content than the control. Most of these results, however, were small in magnitude and confined to the already politically interested Twitter users, as indicated by their pre-treatment tweeting about politics. These findings have implications for social media and news organizations, and also offer direction for future work on how Large Language Models and other computational interventions can effectively enhance individual on-platform engagement with quality news and public affairs.
\end{abstract}

\dates{This manuscript was compiled on \today}

\begin{document}

\dummylabel{sm:fielding-experiment}{1}
\dummylabel{sm:generating-realistic}{2}
\dummylabel{sm:validating-bot-responses}{3}
\dummylabel{sm:outcome-variables}{4}
\dummylabel{sm:account-level-measure}{5}
\dummylabel{sm:pre-and-post-experiment}{6}
\dummylabel{sm:reweighting}{7}
\dummylabel{sm:full-regression}{8}
\dummylabel{sm:alternate-regression}{9}

\maketitle
\thispagestyle{firststyle}
\ifthenelse{\boolean{shortarticle}}{\ifthenelse{\boolean{singlecolumn}}{\abscontentformatted}{\abscontent}}{}

\section{Introduction}
\label{sec:Introduction}
Polarization, declining trust, and wavering support for democratic norms are pressing threats to the U.S. Observers often blame social media platforms for these problems, worrying about echo chambers, misinformation, and algorithmic radicalization \cite{tufekci2018youtube, pariser2011filter, roose2019makingytradicalNYT,hussein2020misinformation}. Evidence to support these worries, however, is limited. Few people inhabit echo chambers \cite{pablo_barbera_social_2020, wojcieszak2022most, fletcher2021many}, encounter or are affected by misinformation \cite{grinberg2019fake, weeks2021, guess2020fake}, or are put in extreme rabbit holes \cite{hosseinmardi2021examining, chen2022subscriptions}. 

We argue that the problem is less that people consume bad political information, but that most people do not consume any at all. News and politics constitute a small fraction of people’s information diets on social media. News makes up only 1.4\% of Facebook’s News Feed \cite{Meta2022, wells2017combining}, the majority of Twitter users do not follow any politicians, journalists, or news organizations \cite{wojcieszak2022most}, and only about 1 in 300 outbound clicks from social media are to substantive news \cite{flaxman2016filter}.\footnote{This is the case online more broadly: only between 2\% \cite{wojcieszak2023no} and 7-9\% \cite{guess2021almost} of all URLs visited by large international samples are news, and - across mobile and desktop - news comprises only 4\% of total consumption \cite{allen2020evaluating}. Although news consumption is greater on television than online, it is still overshadowed by entertainment and other content categories \cite{allen2020evaluating}}). 

This under-consumption of news---and the consequently low levels of political knowledge among the electorate---have important implications \cite{carpini1996americans}. Low-information voters are likely to withdraw from politics, vulnerable to making irrational vote choices \cite{bartels1996uninformed, fording2017cognitive, lau2001advantages}, and easily swung by irrelevant stimuli, emotional appeals, and populist rhetoric in the political environment \cite{achen2006feels}. If sizable, these voters can swing elections \cite{fording2017cognitive, ekins2017five}. In turn, exposure to verified and ideologically balanced quality news creates an informed and efficacious public and leads to more stable political attitudes, lower susceptibility to misinformation, greater acceptance of democratic norms, and voting in accordance with one’s personal and group interests \cite{carpini1996americans, delli2002internet, lupia2000institutional}. Some note that news exposure is the key predictor of political knowledge, seen as ``a demonstrably critical foundation for good citizenship'' \cite{delli2002internet}.

Given these benefits, it is of considerable interest to promote consumption of quality news on social media. Toward this end, this project aims to incentivize Twitter (current X) users to engage with \textit{verified and ideologically balanced} public affairs information.\footnote{We use the old company name, given that the platform was named Twitter during data collection. We focus on Twitter because it is an important channel citizens use to get news and political information \cite{yu2023partisanship} and because it made the field experiment possible at the time of the study \cite{mosleh2022field}.} Specifically, we conduct a large-scale field experiment on 28,457 US-based Twitter users who tweet about lifestyle, entertainment, and sports, topics that dominate social media discourse (see Supplementary Materials [SM] \ref{sm:fielding-experiment} for details on the sample and its selection). We rely on NLP-trained bots to reply contextually and in real time to original non-political tweets of these active users over two weeks. Our GPT-2-generated responses include a relevant reply, specific to the content of the original tweet (e.g., “He’s the best player in the league” as a response to a tweet about a baseball pitcher, see SM \ref{sm:generating-realistic} for examples). In addition, the responses include two core hardcoded elements: a link to topic-relevant non-political section of a verified and ideologically balanced news media organization and an encouragement for the users to follow the Twitter account of that organization. To identify verified and balanced news outlets, we apply validated expert metrics based on human coding from Adfontes media, selecting only the outlets that score high on reliability and low on partisan bias (see SM \ref{sm:validating-bot-responses}A for details on outlet selection and SM \ref{sm:generating-realistic} for the hardcoded elements). Our sample was randomly assigned to either one of two intervention groups, receiving responses from bot accounts presented as either male or female for two weeks, or a control group. We rely on our extensive curated list of U.S. news organizations (see SM \ref{sm:validating-bot-responses}) and a validated BERT-based classifier that identifies users' tweets about politics (see SM \ref{sm:outcome-variables}) to test whether users (a) follow one of the verified and balanced news accounts on our list, (b) retweet content from news media organizations in general, (c) tweet or retweet political content, and (d) like content from news media organizations, and (e) like political content. 

We find that encouragement to follow news through our tailored NLP-based responses had some promising yet limited causal effects; it encouraged the users to follow more news outlets and additionally encouraged those who received comments from the female bot to like more news content on social media. The treatment had no effects on the other outcomes analyzed, i.e. (re)tweeting news content and tweeting about or liking posts about politics. The increases in the liking of news media content were confined to those with high initial levels of political interest, as indicated by previous tweeting about politics, suggesting reinforcement of pre-existing engagement among those already engaged \cite{heissmatthes2019, Nanz_Matthes_2022}. Also, the effects on news media content liking were especially pronounced for the users who were tweeting about sports, with the effects among those who tweeted about entertainment or lifestyle being statistically insignificant due to the decreased sample sizes in these two topic categories.

\begin{figure*}[h]
    \centering
    \includegraphics[width=1.04\textwidth]{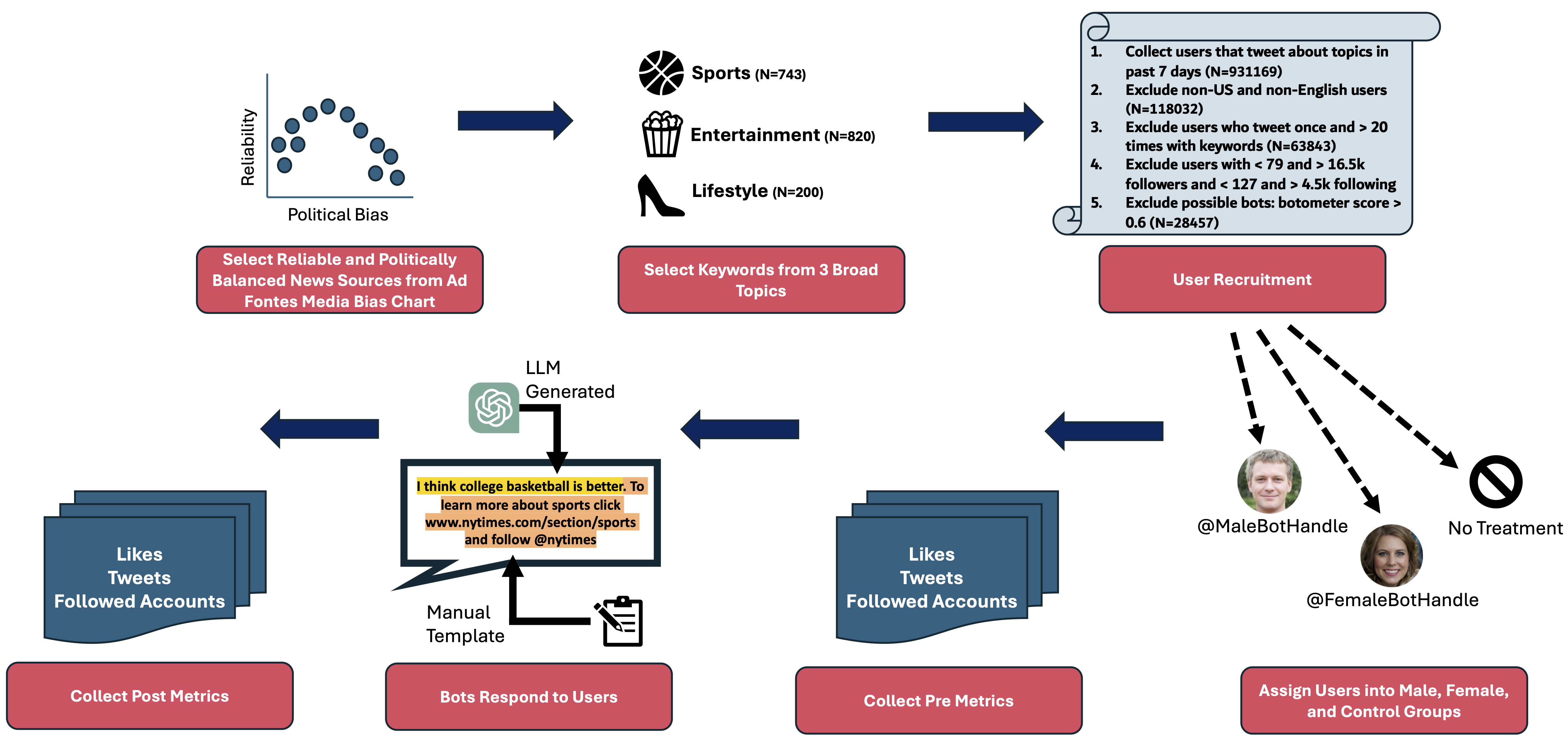}
    \caption{Overview of the Experiment Design.}
    \label{fig:overview}
\end{figure*}

This project advances past work in several key ways. First, we address the core problem of low news use and engagement \cite{Skovsgaard_Andersen_2020, Newman2019, Villi_Aharoni_Tenenboim-Weinblatt_Boczkowski_Hayashi_Mitchelstein_Tanaka_Kligler-Vilenchik_2022}. The overwhelming majority of social media users go online for entertainment, not news or politics \cite{Meta2022, thorson2021algorithmic, eady_how_2019, wojcieszak2020social, Mukurjee2022, mcclain2021behaviors}. Because information exposure on platforms is primarily driven by recommender algorithms that make automated decisions on what content to display based on the user’s past behavior and inferred interests \cite{tiktok21},\footnote{Recommendation algorithms drive between 75-95\% of information consumption on platforms \cite{tiktok21}.}, those users are mostly recommended contents about sports, movies, or celebrities. These personalized recommendations ultimately create closed loops of entertainment consumption and narrow information repertoires \cite{rossi2021closed, thorson2021algorithmic}. Our intervention encourages users' to \textit{follow} news organizations. This puts public affairs information in the users' inventory, thus increasing the likelihood that users have such information readily available in their feed \cite{nyhan2023like}. In addition, following news accounts signals some political interest to the algorithm, thus generating subsequent recommendations to news and political content \cite{thorson2021algorithmic}.

Second, we propose that social media users can be encouraged to consume news and public affairs through their non-political interests. We engage users interested in sports, entertainment, and lifestyle by connecting news with these topics and directing users to news outlets that offer both hard news and softer news about the topics of interest to the users (e.g., coverage of sports, movies, or wellness or style). The work on soft news or ''infotainment'' indeed suggests that programs that discuss cooking or celebrities, but also mention current affairs, attract viewers whose primary motivation is not politics, but who nevertheless learn about current affairs and become more politically active \cite{Andersen2019softnews, Prior2003, baum2006oprah, baek2009don, moy2005communication, Baum2010softnews}. Accordingly, by starting from citizen’s interests in sports, entertainment, or lifestyle topics such as yoga or keto diet, and by directing them to topically relevant content in primarily hard news outlets (e.g., the lifestyle or sports sections of ABC News), we can enhance users' interest and sustainably increase their exposure to quality news. Social media platforms have the potential to act as an intermediary to news media organizations \cite{StierMangoldScharkowBreuer2022, wojcieszak2022avenues} and so encouraging users to follow news accounts and to visit news sites through links embedded in posts may serve as a gateway to hard news consumption \cite{Baum2003}.

Third, differential engagement may occur based on who is sharing news with the users and also who the users are. In general, news and politics are seen as male-dominated spaces \cite{asr2021, Sui2022} and, notably, female media figures are found to receive more toxic, abusive, and hostile responses compared to their male counterparts \cite{Eckert2017, chen2020, Lewis2020}. This is especially true in political contexts \cite{sobieraj2020credible} and this disparity may be most pronounced on social media \cite{usher2018twitter}. It is likely then that users may be less inclined to open links from female sources or to see them as a news source worth following or interacting with. In addition, there may be heterogeneous treatment effects by users' political interest. On the one hand, those politically disinterested individuals may gain more in terms of knowledge, engagement, or subsequent news seeking from inadvertently encountering politics \cite{baum2006oprah, Baum2003}, such as seeing links to news websites on social media platforms. On the other hand, inadvertent exposure to news and politics may reinforce existing gaps in prior political interest \cite{naomi2020fedtos, heissmatthes2019}, such that those who are already interested in news and politics would follow more news accounts and engage most with public affairs online. In short, individuals may be responsive to social media nudges but the source (gender) of these prompts as well as prior users' posting about current affairs on the platform may be important factors influencing these results.

\section{Data and Measurement}
\label{sec:Data}

We identified U.S.-based Twitter users who actively tweeted about one of three topics: sports, entertainment, and lifestyle, across a one week period in September 2022. To do this, we created a list of 1,763 keywords generated using word embeddings and manual additions (e.g., current movies and television series, athletes, brands; see SM \ref{sm:fielding-experiment}A for details; keywords broken down by topic are available at \href{https://github.com/HadiAskari/TwitterBot-Resources/blob/main/latest_keywords.csv}{Github}). We collected our initial user base by scraping the user IDs of all Twitter users who tweeted our keywords at least once in a 7 day period, with location and language filters to ensure that only users based in the U.S. and tweeting in English were included (N = 118,032). We then excluded those who tweeted only once during the 7 day period, as these infrequent users were relatively unlikely to be active during the treatment period. To minimize the chances that power users or administrative accounts (e.g., celebrities, brands, or organizations) are disproportionately represented in our sample, we also excluded users who tweeted more than 20 times (N remaining = 63,843) and those who were in the top 10th and 90th percentiles of followers and followees (i.e., those who had fewer than 79 or more than 16,500 followers and those who followed fewer than 127 or more than 4,500 accounts). Finally, we removed all users with a botometer score \cite{BotometerAPI} of more than 0.60 to minimize the inclusion of bots. This resulted in a final sample of 28,457 active non-bot U.S. users known to tweet about the three topics more than once a week.

These users were randomly assigned to one of three groups: a control, a male bot treatment or a female bot treatment. Randomization was successful on a range of account level metrics (the total number of followed accounts, total number of followers, total posts, and total likes) as well as central pre-treatment metrics (the number of news media accounts followed, the number of recent likes of news media posts, the number of (re)tweets of posts from media accounts), ensuring balance across groups in terms of existing engagement with news media content. All pre-experiment metrics were collected one week prior to the beginning of the experiment using the Twitter API. SM \ref{sm:fielding-experiment}B shows the details on the assignment and randomization. 

We  created 28 bots utilizing GPT-2 to contextually reply to the users in the sample (14 bot accounts for male, and 14 for female treatment group). The bots were designed to be realistic and substantively similar, with gender-definable headshot pictures, gender-identifiable names, and a history of news-related content in their feed (Figure 2 shows two examples, see SM \ref{sm:fielding-experiment}D for additional details on bot creation).\footnote{For consistency and to not introduce additional confounders, all profile pictures of the bots were white.}  

To generate responses to the treated users, we leveraged GPT-2 models \cite{zhang2019dialogpt}.\footnote{At the time of the preparation and execution of the study, higher quality open source models (like Llama2) were not available and it was not feasible to use GPT-3 due to the scale of our experiment and the associated costs.} This model was fine-tuned on Reddit comments by Microsoft and was specifically designed to be conversational in nature. This ensured that the responses were contextually relevant and applicable to the original tweet sent by a user. This contextual nature of the responses, i.e., the fact that each was different and adapted to the original user's tweet, also reduced the likelihood that they were considered spam and banned by Twitter. Before sending the Tweets to the GPT-2 model, we removed all URLs and special characters. Additionally, we discarded the GPT-2 response if (a) it contained language pertaining to Reddit (such as upvote, subreddit, etc), (b) kept on repeating the same text, and (c) used profanity. In cases where responses were discarded, the contextual text was replaced by a randomly selected hardcoded template response. In addition to the GPT-2 based reply to each user's tweet, we hardcoded two elements into the response. We encouraged users to follow a news media organization (e.g. ``follow @wsj'' or ``follow @nyt'') and to visit a link to a relevant sub-section of a verified and ideologically balanced news source (e.g. an entertainment/sports/lifestyle section of the Wall Street Journal or the New York Times). More details on this process can be found in SM \ref{sm:generating-realistic}. 
To ascertain that our intervention directed users to verified and balanced news outlets, we applied validated expert metrics. We compiled a list of reliable and ideologically balanced news sources from Ad Fontes' Media Bias Chart \cite{otero2019ad}. Ad Fontes relies on manually labeled articles, radio, TV, and videos (episodes) from numerous news sources. Each episode is rated by trained human coders and scores are assigned for reliability (from “contains inaccurate/ fabricated information” to “original fact reporting”) and ideological bias (from “most extreme left” to “most extreme right”). We selected sources with a reliability score higher than 40 and a bias score between -18 and 18 (see SM \ref{sm:validating-bot-responses}A for details). These sources, their reliability and bias scores, and the URLs to the relevant sub-sections are presented in SM \ref{sm:validating-bot-responses}A.

The experiment was fielded between 1/19/2023 and 2/3/2023. Every 8 hours we scraped the timelines of all users. Tweets matching one of our topic keywords would then receive an automated reply from an assigned bot account, which contextually and dynamically matched the reply to the original tweet of a user. Each response also encouraged the user to stay up to date with the news and visit a link to a topic-relevant sub-section of a news source from our list, as aforementioned. We limited the number of responses to one per day, so as to ensure that the users who tweet using our topic keywords multiple times a day would not be irritated or seeing our responses as spam. The scraping and response cycle ran continuously for two weeks. After this time period, the treatment to all groups was terminated.  

We collected 3 pre-treatment and post-treatment behavioral user metrics:the followed accounts (pre N = 6,536,692, post N = 17,286,211).\footnote{Followed Account collection using the Twitter API was inconsistent over time, preventing us from collecting all the followed accounts for the entire sample. Specifically, Twitter API's get\_friends() endpoint was returning inconsistent results during the pre-treatment stage. We validated the results several times after the experiment, but during the pre-treatment collection, the API call returned only a subsection of a user's following list. As such, we have full followed accounts data for 11,254 of our users collected at both the pre- and post-experiment stages and we therefore base results for this measure on only these users. Within these collected users, some exhibited unusually high or low decreases in total followers, suggesting either unusual account activity or unresolved inconsistency in API collection results. We therefore only include those whose total followed accounts changed by less than a 50\% increase and more than a 20\% decrease (approximately the 10th and 90th percentiles) and increased by an absolute count of less than 200. We also show robustness checks for alternative cutoffs around these follower numbers in SM \ref{sm:alternate-regression}}, tweets or retweets (pre N = 2,285,401, post N = 2,201,009), and likes among all the users (pre N = 2,927,951, post N 2,846,354).

\begin{figure}[h]
    \centering
    \includegraphics[width=0.5\textwidth]{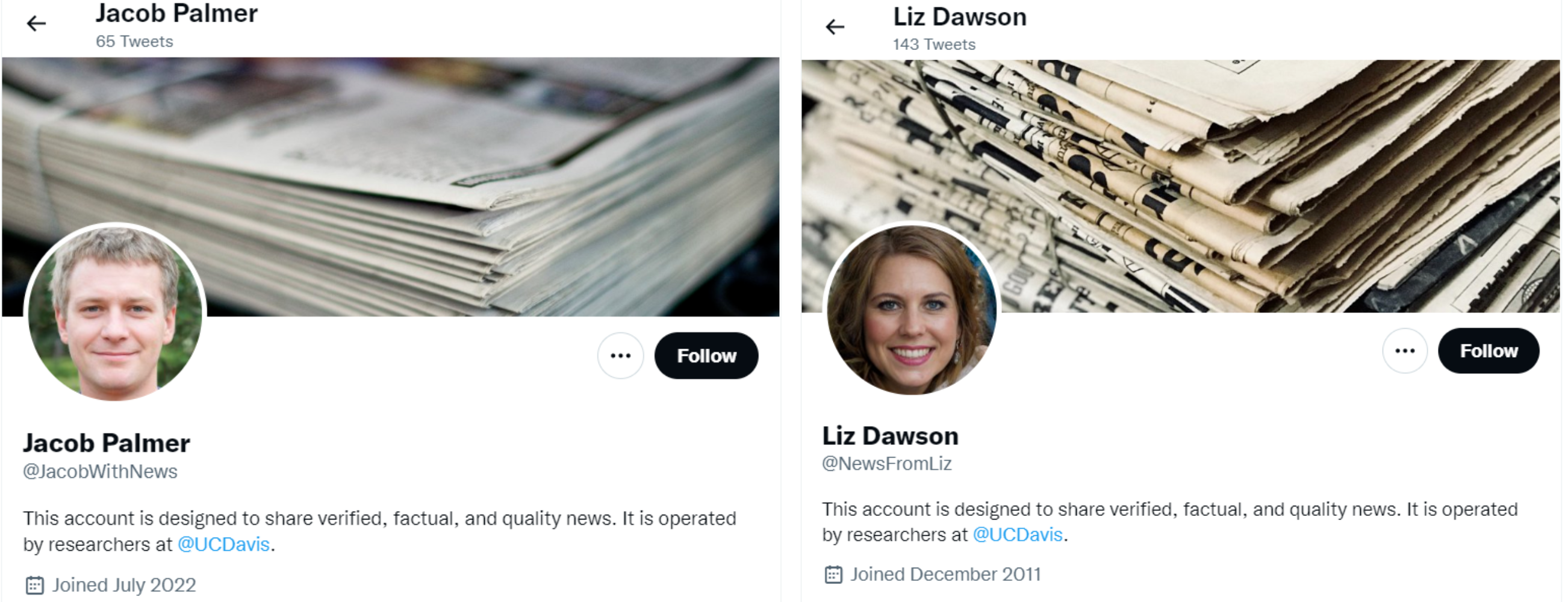}
    \caption{Sample Male and Female Bot Accounts.}
    \label{fig:bots}
\end{figure}


To examine if our intervention increased engagement with news and politics on Twitter, we collected post-treatment metrics one week after the termination of the treatments, contrasting these results to pre-treatment collections of the same measures (based on the prior 100 (re)tweets and likes from a user before the treatment period). We first assessed whether users followed the news organizations from the AdFontes list (which were mentioned in the bot responses to the users) by scraping the accounts each user followed before and after the experiment (with the challenges identified above) to identify any additional news organizations followed. To measure whether users (re)tweeted or liked news content, we used our extensive curated list of over 5,341 News Media organizations. The details on the creation of the overall list are presented in SM \ref{sm:outcome-variables}B and the list is made publicly available on \href{https://github.com/HadiAskari/TwitterBot-Resources/blob/main/media_political_twitter.csv}{Github} 
We identified Twitter handles for 5,341 news organizations from the overall list and identified each user's likes and (re)tweets from these news outlets. 

To accurately measure whether users (re)tweeted or liked political content on Twitter, we developed a fine-tuned BERT-based classifier of political content \cite{BERTPoliticalClassifier}. We conceptualize ``politics'' rather broadly: tweets considered as political include references to political figures, policies, elections, and specific political events \textit{as well as} issues such as climate change, immigration, healthcare, gun control, sexual assault, racial, gender, sexual, ethnic, and religious minorities, the regulation of large tech companies, and crimes involving guns. The classifier was specifically trained on social media data and identifies content about politics with high accuracy (accuracy = .93, precision = .92, recall = .91, F1 = .915). SM \ref{sm:outcome-variables}C shows the details on model training, fine-tuning, and performance. For each user in our sample, we identified all the instances of liking and tweeting political content on the platform.

We also measure to what extent the treated users interacted with our bots, by checking whether they replied to it. Lastly, we evaluate the sentiment of the replies using a RoBERTa-base model trained on 124M tweets from January 2018 to December 2021, and fine-tuned for sentiment analysis with the TweetEval benchmark \cite{loureiro2022timelms,barbieri2020tweeteval}. We collected a total of 241 (99 Male and 142 Female) responses and examined their sentiment. See SM \ref{sm:outcome-variables}E for details.


These measures together comprehensively portray users' pre- and post-treatment posting about and engagement with both news and politics. Treatment effects are examined as the difference in pre- and post-treatment measures for both the female and male treatment groups compared to the control group. For the difference in news media accounts followed, this measure is taken as a simple integer value. For the the difference in the liking and (re)tweeting of news and political contents, these pre- and post-differences are measured as the change in the percentage of (re)tweets and likes from news media accounts or about politics, based on user activity measured specifically in the pre-treatment period and the one-week post post-treatment period. 


\section{Results}
\label{sec:Results}
\subsection{Descriptives}
\label{subsec:Descriptives}

We first describe the pre-treatment following and engagement metrics among our participants to offer a baseline. On average, our users followed 13.71 (Female Treatment group) and 13.69 (Male Treatment group) news accounts prior to our treatment. 
The levels of users' engagement with news content, namely the pre-treatment proportion of likes on and (re)tweets of posts coming from one of the 5,341 news organizations relative to all likes and (re)tweets a user had in the pre-treatment collection period, were very low among our sample.\footnote{We note that this low percentage may be a feature of our design and sampling strategy, such that we selected users based on their frequent posting about sports, entertainment, and lifestyle.} For the liking of news media content, this figure was 0.8\% on average, and for (re)tweeting news media content, this figure was 0.4\%. In short, engagement with news media content was a very infrequent activity. The liking of \textit{political} content was substantially more frequent, likely due to our rather broad conceptualization of what constitutes political content (i.e., not only traditionally hard news such as the election, political parties, the economy, etc., but also social issues, such as race, immigration, abortion, etc.; see SM \ref{sm:outcome-variables}C for the details). Likes on political content constituted just under 12\% of all likes among our users. Similarly, the percentage of (re)tweets from our users which were related to politics was around 11.5\% across user groups. 

\begin{figure*}[h]
    \centering
    \includegraphics[width=\textwidth]{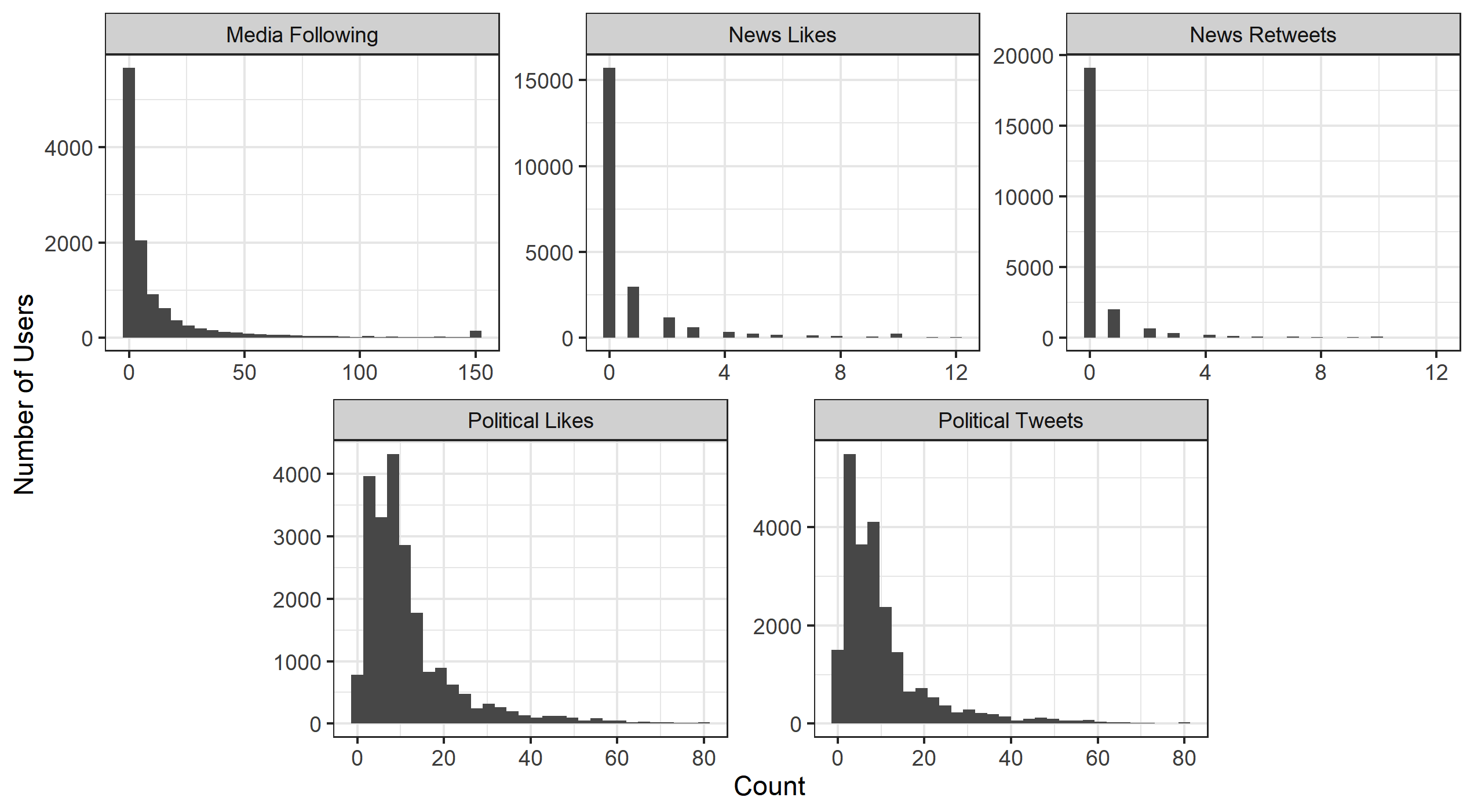}
    \caption{User Distribution Across Pre-Treatment Measures. Followed news media accounts are a count measure based on all recorded accounts followed. News media likes, news media (re)tweets, political likes, and political (re)tweets are measured as a count based on the last 100 likes or (re)tweets made prior to the treatment period.}
    \label{fig:barplots}
\end{figure*}

Looking at the distributions of these variables in Figure \ref{fig:barplots}, it can be seen that many users do not follow any news media accounts and do not engage with any content from news media organizations on Twitter. At the same time, the vast majority of users in our sample do like political content and (re)tweet about politics in some form, with most users doing so in between 5\% and 20\% of their likes and (re)tweets, respectively. At the aggregate level, then, the results suggest relatively limited news engagement, as consistent with prior work \cite{Meta2022, wojcieszak2022most}, and greater engagement with political content.

In terms of user activity during our treatment period, across the two week intervention period, our users (re)tweeted a total of 1,172,143 (re)tweets (396,378 by users in the male bot treatment group, 367,672 by users in the female bot treatment group, 408,093 by users in the Control), with 154,878 (13.2\%) of those containing text that matched one or more words in our keyword list. Of these matches, 76.67\% of (re)tweets were related to the topic of sports, 17.56\% related to entertainment and 5.78\% to lifestyle. Based on our self imposed limit of one response per user per 24 hours, our bots then responded to 28,211 of these (re)tweets.


\subsection{Treatment Effects}
\label{subsec:MainEffects}

\begin{figure*}[h]
    \centering
    \includegraphics[width=\textwidth]{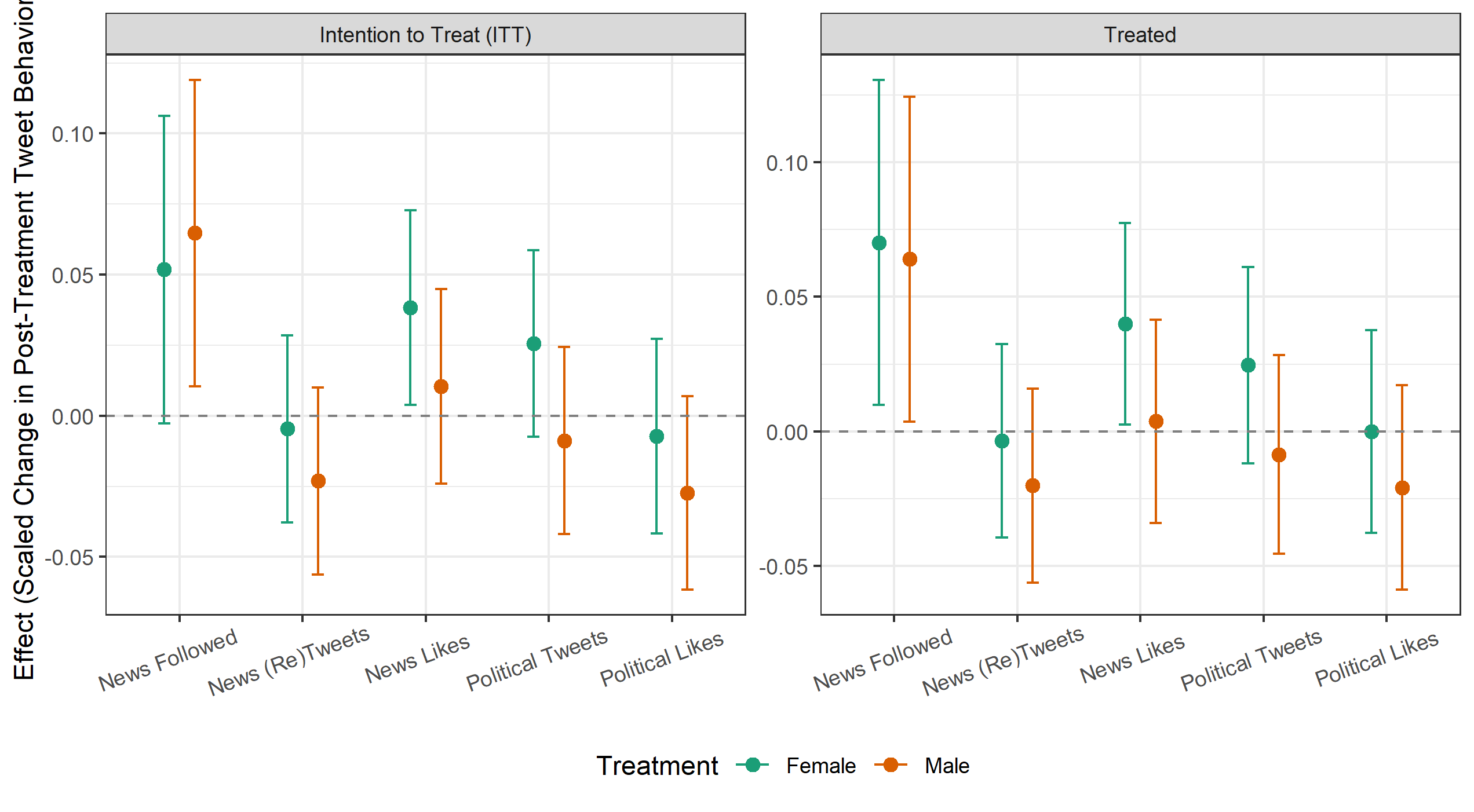}
    \caption{Main Effects Plot: Coefficient estimates and 95\% confidence intervals for G-computation after entropy balancing regression models with robust standard errors. Dependent variables taken as the difference between pre- and post-treatment individual user measures. News media accounts followed measured as a count, news media (re)tweets and likes and political (re)tweets and likes measured as percentages.}
    \label{fig:Main Effects}
\end{figure*}

Modelling the pre- to post-treatment changes in user activity, Figure \ref{fig:Main Effects} shows the estimated treatment effects based on a linear regression model measuring the difference in pre- and post-experiment metrics at the user level, with results shown for the male and female treatment groups compared to the control group. The full models are reported in SM \ref{sm:full-regression}. In the models, the number of news media following is measured as a continuous change in the number of news media accounts followed while news media likes and (re)tweets and political likes and (re)tweets variables are measured as a relative change in the percentage of each measure between pre- and post-treatment.

As our intervention was dependent on users actually tweeting about our keywords during the experimental time period, not all users in the treatment groups received the treatment from one of our bots. To account for this, we estimate two distinct treatment effects: ``Intention to treat (ITT)'' (i.e., all users from our original randomized treatment groups) and ``Treated'' (i.e., users in the male and female treatment groups who actually received one or more responses from our bots). Because certain users may have been more likely to tweet matching keywords and view the bot responses, and so simply dropping untreated units would result in imbalance between the control group and the refined treatment groups, we use an entropy balancing approach \cite{hainmueller_2012} to reweigh our ``treated'' treatment groups relative to the control group. The entropy balancing is based on four account level metrics that capture the overall size and activity levels of an account and that are also correlated with the likelihood of receiving and seeing a treatment (total likes, total tweets, total followers, and total followed accounts). After balancing, the estimates are computed using G-computation \cite{chattonetal,robins1986} with robust standard errors. For comparability, estimates are standardized with effects interpreted in standard deviation changes.

Figure \ref{fig:Main Effects} shows that, in our ITT models, users in the female treatment group liked significantly more content from news media accounts compared to the control group, and that those in the male bot treatment group liked significantly more content from news media organizations. The other variables showed no statistically significant change in the ITT models. When examining the users who were indeed treated with bot responses during the experimental period (see the ``Treated'' models in Figure \ref{fig:Main Effects}), we see that users in both the female and male bot treatment groups were significantly more likely to follow news media accounts (p=0.05) than those in the control. Each user in the female bot treatment group followed, on average, 0.75 more news media accounts than the control, and those in the male bot treatment followed 0.69 more news media accounts during the treatment period. In addition, those in the female bot treatment group liked significantly more news media content (with a consistent coefficient of a 0.04 standard deviation increase). Given that the median number of pre-treatment following and liking of news media was 0, these effects are meaningful. Nevertheless, although meaningful and significant, these effects were substantively very small. 
Also, we find no statistically significant effects of our intervention on the three remaining outcomes: (re)tweeting tweets from news media accounts, (re)tweeting political content, and liking political content.

In general, then, among those users who actually received our treatments, users followed slightly more news media accounts and those who were treated by a female bot additionally liked more news media content, suggesting small differential effects based on the gender of the bot. 

\begin{figure*}[h]
    \centering
    \includegraphics[width=\textwidth]{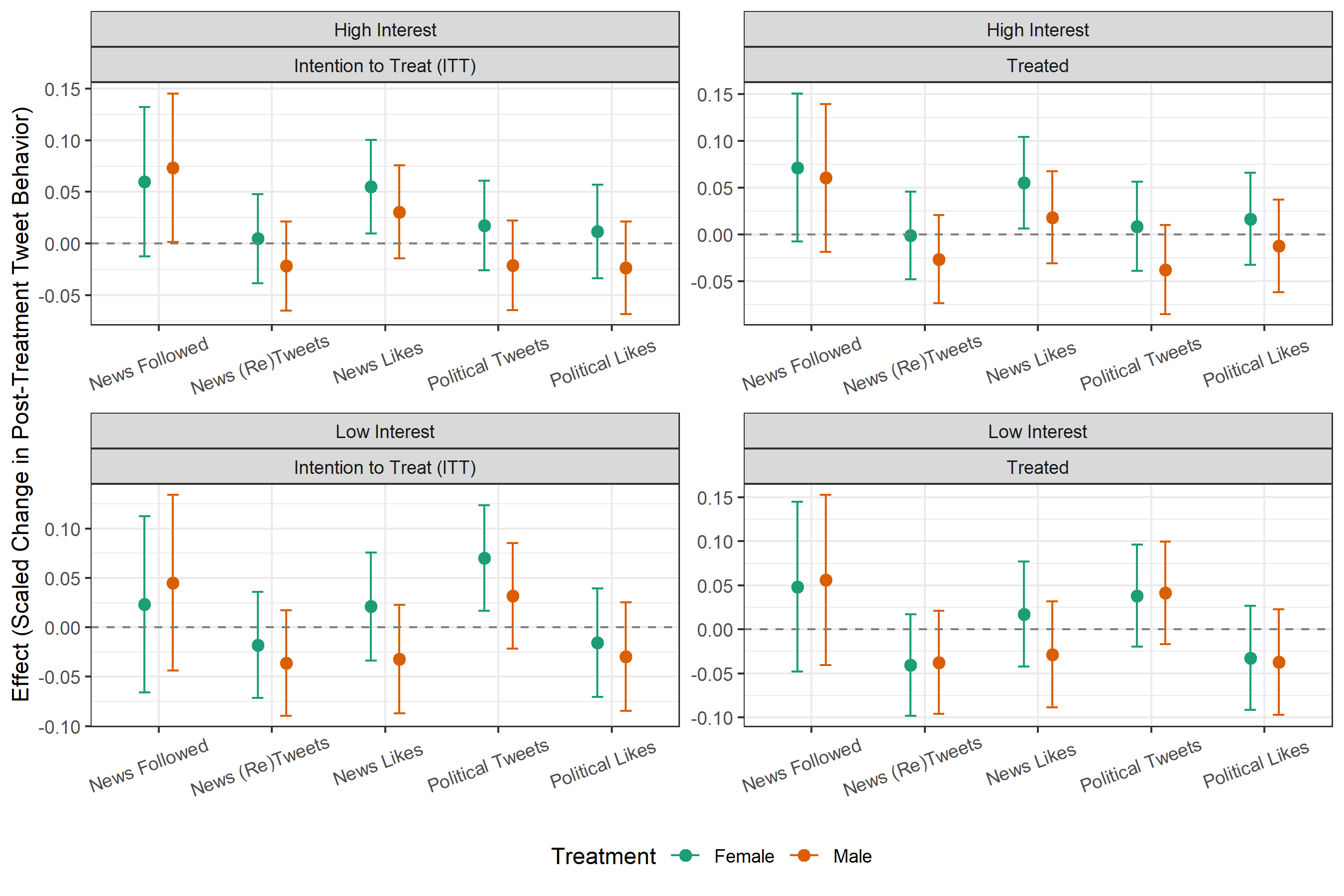}
    \caption{Main Treatment Effects Divided by Users' Prior Political Engagement Levels:Coefficient estimates and 95\% confidence intervals for G-computation after entropy balancing regression models with robust standard errors. Dependent variables taken as the difference between pre- and post-treatment individual user measures. News media accounts followed measured as a count, news media (re)tweets and likes and political (re)tweets and likes measured as percentages.}
    \label{fig:PolInterest Effects}
\end{figure*}

Given that the average engagement with news and political content was relatively low, the question arises as to which types of users may have been affected by our treatments. Specifically, were the detected increases concentrated among those who were politically engaged already or were the treatments able to trigger some baseline engagement among users who were previously not interested in news and politics? We examine the heterogeneity of the effects by users' prior on-platform engagement with political content (a binary indicator of whether a user (re)tweeted 5 or fewer vs. more times about politics in the pre-treatment period). Figure \ref{fig:PolInterest Effects} shows that the results are only significant in the high political interest group. For those users, the responses from female bots significantly---but again only slightly---increased the liking of news media content. The following of media accounts also increased (although due to the reduced sub-sample size, the p value falls to 0.11, despite the estimated standardized effect size actually increasing relative to the overall model). In the low interest sub-group, no effects in the treated model estimations are significant or approaching significance.

Lastly, we explore whether the treatment effects differed by topic category by running separate models for users who---during the treatment period---tweeted about entertainment versus lifestyle versus sports. We perform topic-wise classification of our users into these 3 categories by doing a combination of keyword matching and BERT-based embedding matching on all of the tweets we collected for each user in our 2 week period. Whichever category received the most tweets by the user was classified as the category for that user. The final user category percentages via this method was 61.30\% Sports, 30.67\% Entertainment and 8.03\% Lifestyle. 

\begin{figure*}[h]
    \centering
    \includegraphics[width=\textwidth]{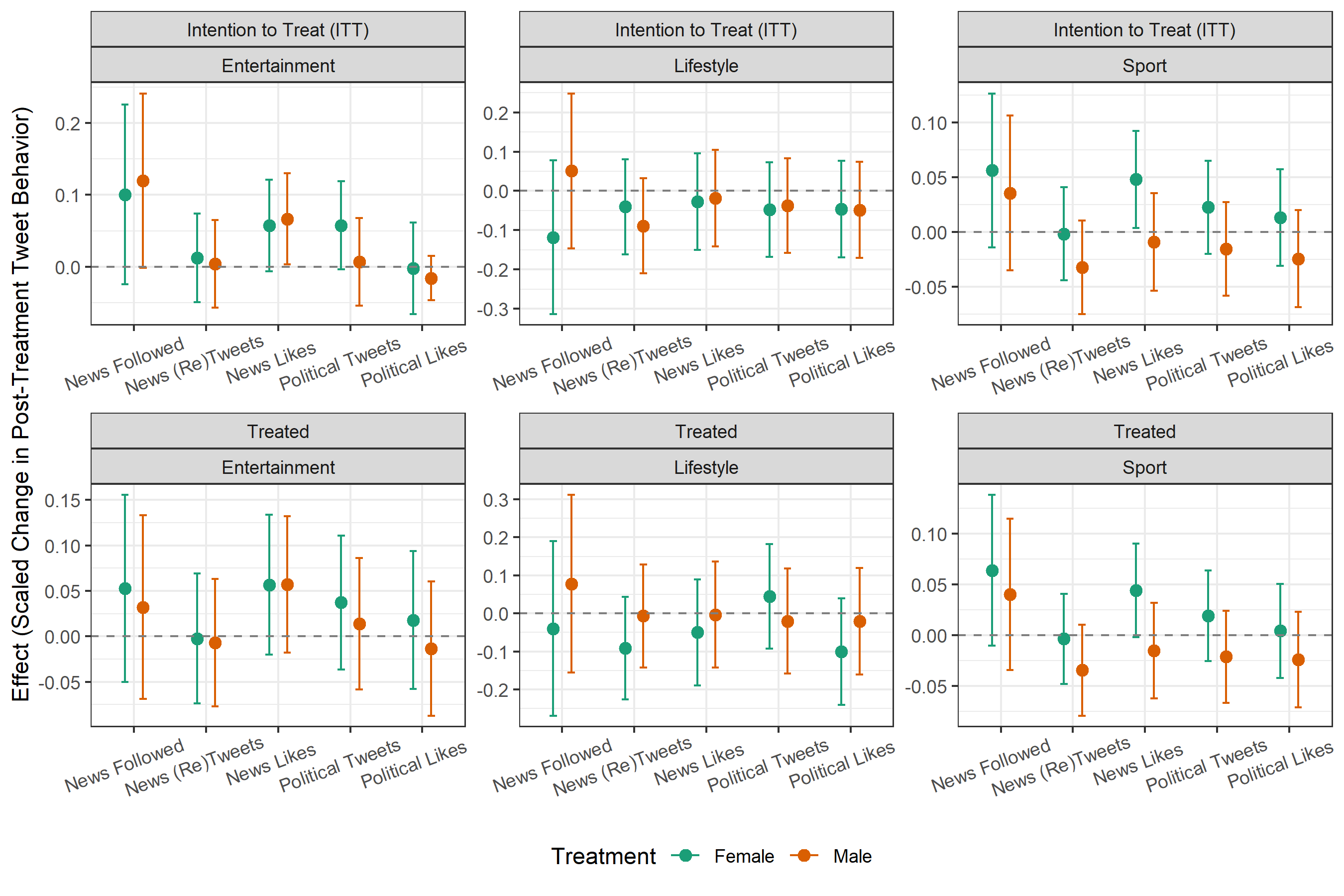}
    \caption{Main Treatment Effects Divided by Users' Primary Topic of Twitter Activity: Coefficient estimates and 95\% confidence intervals for G-computation after entropy balancing regression models with robust standard errors. Dependent variables taken as the difference between pre- and post-treatment individual user measures. News media accounts followed measured as a count, news media tweets and likes and political tweets and likes measured as percentages.}
    \label{fig:Topic Effects}
\end{figure*}

As shown in Figure \ref{fig:Topic Effects}, the effects are very limited, again due to the substantially smaller sub-sample sizes, especially among the treated groups. The only statistically significant effect emerges among users who tweeted about sports. There, we find that the users in the female bot treatment group increased their liking of news media content relative to the control in both the ITT and the treated models. We also see a significant result for the male treatment among the users tweeting about entertainment, finding that comments from a male bot increased the liking of news media content. This effect, however, becomes insignificant once we focus on the treated group.  

\section{Discussion}

The American public is largely disinterested in politics \cite{krupnikov2022other} and so the aggregate consumption of news and political information is limited offline \cite{Skovsgaard_Andersen_2020, Newman2019, Villi_Aharoni_Tenenboim-Weinblatt_Boczkowski_Hayashi_Mitchelstein_Tanaka_Kligler-Vilenchik_2022, allen2020evaluating}, online \cite{wojcieszak2024non} and on social media platforms \cite{Meta2022,wojcieszak2021no, thorson2021algorithmic, nyhan2023like}. And yet, because news consumption has many beneficial effects---ranging from increased political knowledge and participation, to more stable political attitudes, greater political tolerance, and higher support for democratic norms \cite{carpini1996americans, delli2002internet, lupia1998democratic}---the question as to how to incentivize social media users to consume more quality news is important. Because most citizens go online for entertainment \cite{mcclain2021behaviors}, this project proposed to link users' online habits and interests with contextual nudges that could encourage them to consume more news and public affairs information on Twitter. 

Our experiment ran for two weeks on a sample of 28,457 U.S.-based Twitter users interested in sports, entertainment, and lifestyle. To engage those users, we deployed 28 male and female bots trained to contextually respond to the users in online conversations and to suggest topic-relevant sub-pages of credible and ideologically balanced news sources, as determined by external metrics. As suggested by the literature on ``soft news,'' this particular intervention should encourage the users to follow and engage with these sources on Twitter, put more public affairs information in their feed, signal to the platform algorithms that the users are interested in news and politics, and eventually lead the users to consume more ``hard news.'' This intervention offers three key findings.

First, our bot-based intervention, and in particular female bots responding to Twitter users with encouragement to follow credible, ideologically balanced news accounts and to visit topic-specific sub-pages of news organizations led to very minimal changes in news and political engagement on Twitter. This suggests that mere bot-based intervention is an insufficient encouragement to change individual media engagement behavior in any pronounced or meaningful way. As people tend to be resistant to change their habits and as media consumption is a habitual behavior, the finding that our intervention did not increase such aspects of news and political engagement as (re)tweeting content from news accounts, (re)tweeting political content, or liking political tweets may not come as a surprise. That said, our treatments did slightly increase some aspects of news engagement among the treated users. Those users did follow more news accounts and those who were contacted by female bots also liked more posts from news organizations on Twitter. These effects were small in magnitude, producing a less than 0.1 standard deviation change. Although small, these effects \textit{could} be meaningful in ways that cannot be quantified in this project. In particular, the additional following of news accounts puts more content from these accounts in users' social media feed. Exposure to such content could over time increase political knowledge and efficacy, generate engagement with the content seen, and serve as a gateway to following additional accounts and consuming public affairs elsewhere. In addition, this increased following and the liking of news content could send the signal to the algorithm that the user is --- at least to some extent --- interested in public affairs, thereby promoting future recommendations to relevant content and suggestions to follow news or political accounts \cite{haroon}. These subtle and potentially cumulative effects cannot be tested in this project, and so we invite future work to examine them using longer designs that capture additional variables, both behavioral and self-reported. We also encourage researchers to explore how bots could be designed to encourage people to follow quality accounts and click on relevant links more effectively.

Second, the bots presented as females led to more consistent and stronger increases in news following and the liking of news media content. Despite the fact that politics is still seen as a male domain \cite{asr2021, Sui2022} and despite the mounting evidence that women are more likely than men to experience harassment or hate speech on social media \cite{Eckert2017, chen2020, Lewis2020}, our sample was slightly more responsive to the intervention when it came from a female. Because Twitter users are predominantly male (61.2\% vs 38.8\%) \cite{twitterdemo} and because our sample was likely to have even more male users (given that sports was the dominant category), it is possible that men are more open to female bots nudging them and could feel potentially threatened by male bots telling them to follow news. Again, because these effects were small and not very robust, and because this evidence goes against the prevalent finding that politics and the online sphere are spaces where female opinion is disregarded, we encourage more work exploring these potentially differential reactions to pro-social online interventions coming from females versus males. 

Third, the small detected effects were largely confined to the group of users who were already interested in politics. It is those who had previously tweeted about politics who showed the most substantial increases in the liking of news media contents as a result of our intervention (among those who were treated). In contrast, the bots failed to encourage users who were not interested in politics, as determined by their previous posting of political content on Twitter, to follow more news accounts, (re)tweet politics, or like news and political content.
This finding speaks to the reinforcement hypothesis \cite{heissmatthes2019, Nanz_Matthes_2022} and suggests that people who are already politically interested are the ones who become yet more politically interested as a result of various interventions. Although incidental exposure to news and public affairs on social media platforms---which the comments from our bots effectively created---could serve an equalizing function and pull the previously disinterested citizens back into news and politics \cite{baum2006oprah}---our study adds to the more pessimistic evidence that individuals with high political interest are becoming information richer and more participatory, whereas the ones with low political interest remain politically disengaged \cite{heissmatthes2019}. We also note that whatever significant effects mostly emerged among the users who were tweeting about sports, potentially due to the fact that sports users were the majority of our sample. We caution against putting too much leverage on these findings because the sample sizes became substantially smaller when sub-divided into the three topic categories.  

Naturally, the experiment is not free from limitations that offer important directions for future research. It is possible that stronger effects would emerge if we used GPT-4 or other Large Language Models more powerful than our fine-tuned GPT-2. GPT-4 generated replies to users would have likely been even more human-like and potentially better aligned with the original users' tweets, which may have generated stronger effects. In addition, although the experiment ran for 2 weeks, a time frame that is rather extensive, more rounds of bot interactions with our users over an even longer time period could have generated stronger effects. Because many of the users tweeted relatively infrequently (6,477 users out of 28,457 did not tweet even once during the experiment period) and many (3,674 users out of the remaining 21,980) did not tweet our keywords at all during the treatment period, the treatment may have been too weak to generate effects. That said, the bot responses were capped at one per day to avoid spamming or angering the users. Future studies should examine what ``dose'' of various social media interventions is most effective. 

In addition, it is not certain whether similar effects would emerge on a different platform, in a different time, or in a different sociopolitical context. Twitter is known to be an important channel for political information \cite{Pew2022, Pew2021}, a key platform for politicians, journalists, and pundits \cite{Pew2020}, and one where many users express their political opinions and engage in political activities \cite{mcgregor2019social, barbera2019leads}. As such, engaging users with news and politics may have been more ``natural'' on and better integrated with Twitter than with Facebook or Instagram, where many users have more closely knit networks of friends and family. We encourage scholars to replicate our results on other platforms. Given the changes to the Twitter API after Elon Musk acquired Twitter, field experiments such as ours and others \cite{mosleh2021perverse, mosleh2021shared, mosleh2022field} might be no longer possible. 

Lastly, our core focus was on encouraging the users to follow news accounts and engage with news and political information on social media, and so we cannot ascertain whether the treatment or the slightly increased news following and liking of news content had any effects on users' political attitudes. A growing body of evidence suggest that although various (algorithmic) interventions can powerfully alter users' on-platform exposures and behaviors, this has no corresponding effects on affective polarization, misperceptions, policy positions, among other attitudes \cite{nyhan2023like,ventura2023whatsapp, casas2022exposure, guess2023social, haroon}. As aforementioned, however, our treatment could have triggered or enhanced political interest among some users, served as a gateway to hard news, or made users feel politically efficacious, outcomes that we did not measure and that are often overlooked in similar work.

Despite their constrained nature and limited size, our findings have implications for research, platforms, and democracy more broadly. As most social media users do not see or engage with public affairs information on platforms \cite{Meta2022, wojcieszak2022most} in part because they do not have such information in their social media inventory \cite{WellsThorson2018}, scholars should explore ways and design (algorithmic) interventions that encourage citizen engagement with news and politics and that make such content easily available to the users. Although social media researchers disproportionately focus on the (hot and sexy) misinformation and ``echo chambers,'' these digital problems are relevant to a much smaller subset of the population than the low levels of consumption of quality political information. As such, a shift in focus is needed.

In addition, the fact that enhancing the accessibility of credible and ideologically balanced news at least slightly encourages some aspects of citizen news engagement suggests that social media platforms could (and should) introduce such pro-social interventions toward increasing citizen awareness of public affairs. Naturally, platforms prioritize user engagement over the quality or veracity of information \cite{brown2021twitter}. Yet --- as other research shows \cite{haroon}--- algorithmic nudges that increase recommendations and exposure to ideologically balanced quality news do not decrease user engagement on social media platforms. Naturally, more research is needed on how to minimize harmful recommendations and exposures and enhance user engagement with verified public interest content on social media. As most people post about such seemingly non-political issues as yoga, baseball, or a recent movie release, connecting those interests to public affairs holds some promise. Given the widespread use of platforms for content exposure and the various challenges faced by the United States and other democracies, such research is timely and needed.

\matmethods{Our experiment was fielded in 1/19/2023 to 2/3/2023 with data collection and processing occurring in January-February 2023. Collectively, across this period, there were five stages for the setup and execution of our experiment. Firstly, we identified keywords across three distinct popular non-hard-news topic areas, we then collected our user sample for the experiment, followed by their pre-treatment Twitter information, we then ran our news bot intervention on their relevant tweets, and, finally, we collected their post-treatment data. We leverage Tweepy \cite{Tweepy} and several Twitter v1.1 API tokens to perform all experimentation. Details on which API calls we used to conduct the experiment can be found in SM \ref{sm:outcome-variables}D. We now expand on the process below:

We identified U.S.-based Twitter users who actively tweeted about one of three topics: sports, entertainment, and lifestyle, across a one week period in September 2022. To do this, we created a list of 1,763 keywords generated using word embeddings and manual additions (e.g., current movies and television series, athletes, brands; see SM \ref{sm:fielding-experiment}A for details; keywords broken down by topic are available at \href{https://github.com/HadiAskari/TwitterBot-Resources/blob/main/latest_keywords.csv}{Github}). We collected our initial user base by scraping the user IDs of all Twitter users who tweeted our keywords at least once in a 7 day period (using API.search\_tweets()), with location and language filters to ensure that only users based in the U.S. and tweeting in English were included (N = 118,032). We used the package geostring \cite{Geostring} and sPaCy's Language Detector \cite{SpacyLang} to filter location and language respectively. We then excluded those who tweeted only once during the 7 day period, as these infrequent users were relatively unlikely to be active during the treatment period. To minimize the chances that power users or administrative accounts (e.g., celebrities, brands, or organizations) are disproportionately represented in our sample, we also excluded users who tweeted more than 20 times (N remaining = 63,843) and those who were in the top 10th and 90th percentiles of followers and followees (i.e., those who had fewer than 79 or more than 16,500 followers and those who followed fewer than 127 or more than 4,500 accounts). Finally, we removed all users with a botometer score \cite{BotometerAPI} of more than 0.60 to minimize the inclusion of bots. This resulted in a final sample of 28,457 active non-bot U.S. users known to tweet about the three topics more than once a week. More details available in SM \ref{sm:fielding-experiment}B.

Having identified our user pool, we then assigned these users to one of three treatments (a male bot, a female bot, and a control group). Randomization was successful on a range of account level metrics (the total number of followed accounts, total number of followers, total posts, and total likes) as well as central pre-treatment metrics (the number of news media accounts followed, the number of recent likes of news media posts, the number of (re)tweets of posts from media accounts), ensuring balance across groups in terms of existing engagement with news media content. To account for the volume of messaging required, we created multiple bot accounts per treatment group (14 Male and 14 Female). These bots were designed to be realistic at a visual level, with each bot having a clearly gender definable headshot picture and a clearly gender identifiable name. In order to better comply with Twitter's Terms of Service, we included the following in the bio of the accounts “This account is designed to share verified, factual, and quality news. It is operated by researchers @ University of California, Davis”. See SM \ref{sm:fielding-experiment}D for more details which websites we used to create the accounts.  

To generate responses to the treated users, we leveraged GPT-2 models \cite{zhang2019dialogpt}. This model was fine-tuned on Reddit comments by Microsoft and was designed to be conversational in nature. 
Before sending the Tweets to the GPT-2 model, we removed all URLs and special characters and discarded the GPT-2 response if it contained language pertaining to Reddit (such as upvote, subreddit, etc), kept on repeating the same text, or used profanity. In cases where responses were discarded, the contextual text was replaced by a randomly selected hardcoded template response. In addition to the GPT-2 based reply to each user's tweet, we hardcoded two elements into the response. We encouraged users to follow a news media organization (e.g. ``follow @wsj'' or ``follow @nyt'') and to visit a link to a relevant sub-section of a verified and ideologically balanced news source (e.g. an entertainment/sports/lifestyle section of the Wall Street Journal or the New York Times). More details on this process can be found in SM \ref{sm:generating-realistic}. As mentioned in the main text, the bot responses were encouraging users to follow and visit news media organizations that have been established to be verified and ideologically balanced. All the sources, their reliability and bias scores, and the URLs to the relevant sub-sections are presented in SM \ref{sm:validating-bot-responses}A.

%

Every 8 hours we scraped the timelines of all users using API.user\_timeline(). Tweets matching one of our topic keywords would then receive an automated reply from an assigned bot account, which contextually and dynamically matched the reply to the original tweet of a user using API.update\_status(). Each response also encouraged the user to stay up to date with the news and visit a link to a topic-relevant sub-section of a news source from our list, as aforementioned. We limited the number of responses to one per day, so as to ensure that the users who tweet using our topic keywords multiple times a day would not be irritated or seeing our responses as spam. The scraping and response cycle ran continuously for two weeks. After this time period, the treatment to all groups was terminated.





\subsection*{Outcome measurement}
We measure three different variables across conditions pre- and post-intervention, namely (1) how much (a) political content (b) content from news organizations users (re)tweet (i.e., tweet, retweet and quote tweet), (2) how much (a) political content (b) content by news organizations they like, (3) how many news accounts they follow. To do so, we collected the following for all subjects \textit{before} the intervention: (1) their last hundred (re)tweets before the start of the experiment, which we classify as (a) political or not with a BERT classifier; and which we (b) categorize as coming from a news/political/media account or not (based on an extensive list of 5400 US news organizations and 5,341 Twitter handles, as well as a list of political/media personalities) \href{https://github.com/HadiAskari/TwitterBot-Resources/blob/main/media_political_twitter.csv}{Github} (2) the last hundred ``likes'', which we classify as being on content that is (a) political or not with a BERT classifier; and on content which we (b) categorize as coming from a news/political/media account or not  (based on the same list); (3) the list of accounts they followed at the start of the experiment, which we use to determine the number of news/political/media accounts followed (based on the same list). \textit{After} the experimental manipulation, we collected equivalent variables. 

The API call we used to get the followed accounts was API.get\_friends\_ids(). The call for likes was API.get\_favorites() and the call for (re-)tweets was API.user\_timeline(). The final counts collected were as follows: followed accounts (pre N = 6,536,692, post N = 17,286,211), (re-)tweets (pre N = 2,285,401, post N = 2,201,009), and likes (pre N = 2,927,951, post N 2,846,354). More details can be found in SM. \ref{sm:outcome-variables}




}

\showmatmethods{} 

\acknow{The authors gratefully acknowledge the support of the European Research Council, “Europeans exposed to dissimilar views in the media: investigating backfire effects,” Proposal EXPO- 756301 (ERC Starting Grant, Magdalena Wojcieszak -- PI). The authors are also grateful to Zubair Shafiq and Muhammad Haroon for their research assistance. Any opinions, findings, and conclusions or recommendations expressed in this material are those of the authors and do not necessarily reflect the views of the European Research Council.}

\showacknow{} 


\bibliography{refs_FINAL}

\begin{thebibliography}{10}

\bibitem{tufekci2018youtube}
Z Tufekci, Youtube, the great radicalizer.
\newblock {\em\protect\JournalTitle{The New York Times}} \textbf{10}, 2018
  (2018).

\bibitem{pariser2011filter}
E Pariser, {\em {The Filter Bubble: What the Internet Is Hiding from You}}.
\newblock (Penguin Press), (2011).

\bibitem{roose2019makingytradicalNYT}
K Roose, {The Making of a YouTube Radical. The New York Times}
  (\url{https://www.nytimes.com/interactive/2019/06/08/technology/youtube-radical.html})
  (2019).

\bibitem{hussein2020misinformation}
E Hussein, P Juneja, T Mitra, {Measuring Misinformation in Video Search
  Platforms: An Audit Study on YouTube} in {\em ACM Conference On
  Computer-Supported Cooperative Work And Social Computing (CSCW)}.
\newblock (2020).

\bibitem{pablo_barbera_social_2020}
{Pablo Barbera}, Social {Media}, {Echo} {Chambers}, and {Political}
  {Polarization} in {\em Social {Media} and {Democracy}: {The} {State} of the
  {Field}, {Prospects} for {Reform}}, eds.{} N Persily, JA Tucker.
\newblock (Cambridge University Press, Cambridge), (2020).

\bibitem{wojcieszak2022most}
M Wojcieszak, A Casas, X Yu, J Nagler, JA Tucker, Most users do not follow
  political elites on twitter; those who do show overwhelming preferences for
  ideological congruity.
\newblock {\em\protect\JournalTitle{Science advances}} \textbf{8}, eabn9418
  (2022).

\bibitem{fletcher2021many}
R Fletcher, CT Robertson, RK Nielsen, How many people live in politically
  partisan online news echo chambers in different countries?
\newblock {\em\protect\JournalTitle{Journal of Quantitative Description:
  Digital Media}} \textbf{1} (2021).

\bibitem{grinberg2019fake}
N Grinberg, K Joseph, L Friedland, B Swire-Thompson, D Lazer, Fake news on
  twitter during the 2016 us presidential election.
\newblock {\em\protect\JournalTitle{Science}} \textbf{363}, 374--378 (2019).

\bibitem{weeks2021}
BE Weeks, E Menchen-Trevino, C Calabrese, A Casas, M Wojcieszak, Partisan
  media, untrustworthy news sites, and political misperceptions.
\newblock {\em\protect\JournalTitle{New Media and Society}} \textbf{0} (2021).

\bibitem{guess2020fake}
AM Guess, et~al., `fake news' may have limited effects beyond increasing
  beliefs in false claims.
\newblock {\em\protect\JournalTitle{Harvard Kennedy School Misinformation
  Review}} \textbf{1} (2020).

\bibitem{hosseinmardi2021examining}
H Hosseinmardi, et~al., Examining the consumption of radical content on
  youtube.
\newblock {\em\protect\JournalTitle{Proceedings of the National Academy of
  Sciences}} \textbf{118}, e2101967118 (2021).

\bibitem{chen2022subscriptions}
AY Chen, B Nyhan, J Reifler, RE Robertson, C Wilson, Subscriptions and external
  links help drive resentful users to alternative and extremist youtube
  channels.
\newblock {\em\protect\JournalTitle{Science Advances}} \textbf{9}, eadd8080
  (2023).

\bibitem{Meta2022}
Meta, Widely viewed content report: What people see on facebook (2022)
  \url{https://transparency.fb.com/data/widely-viewed-content-report/#intro}.

\bibitem{wells2017combining}
C Wells, K Thorson, Combining big data and survey techniques to model effects
  of political content flows in facebook.
\newblock {\em\protect\JournalTitle{Social Science Computer Review}}
  \textbf{35}, 33--52 (2017).

\bibitem{flaxman2016filter}
S Flaxman, S Goel, JM Rao, Filter bubbles, echo chambers, and online news
  consumption.
\newblock {\em\protect\JournalTitle{Public opinion quarterly}} \textbf{80},
  298--320 (2016).

\bibitem{wojcieszak2023no}
M Wojcieszak, et~al., No polarization from partisan news: Over-time evidence
  from trace data.
\newblock {\em\protect\JournalTitle{The International Journal of
  Press/Politics}} \textbf{28}, 601--626 (2023).

\bibitem{guess2021almost}
AM Guess, (almost) everything in moderation: New evidence on americans' online
  media diets.
\newblock {\em\protect\JournalTitle{American Journal of Political Science}}
  \textbf{65}, 1007--1022 (2021).

\bibitem{allen2020evaluating}
J Allen, B Howland, M Mobius, D Rothschild, DJ Watts, Evaluating the fake news
  problem at the scale of the information ecosystem.
\newblock {\em\protect\JournalTitle{Science Advances}} \textbf{6}, 1--6 (2020).

\bibitem{carpini1996americans}
MXD Carpini, S Keeter, {\em What Americans know about politics and why it
  matters}.
\newblock (Yale University Press), (1996).

\bibitem{bartels1996uninformed}
LM Bartels, Uninformed votes: Information effects in presidential elections.
\newblock {\em\protect\JournalTitle{American journal of political science}} pp.
  194--230 (1996).

\bibitem{fording2017cognitive}
RC Fording, SF Schram, The cognitive and emotional sources of trump support:
  The case of low-information voters.
\newblock {\em\protect\JournalTitle{New Political Science}} \textbf{39},
  670--686 (2017).

\bibitem{lau2001advantages}
RR Lau, DP Redlawsk, Advantages and disadvantages of cognitive heuristics in
  political decision making.
\newblock {\em\protect\JournalTitle{American journal of political science}} pp.
  951--971 (2001).

\bibitem{achen2006feels}
CH Achen, LM Bartels, It feels like we’re thinking: The rationalizing voter
  and electoral democracy in {\em Annual Meeting of the American Political
  Science Association, Philadelphia}.
\newblock Vol.{}~30, p.~1 (2006).

\bibitem{ekins2017five}
E Ekins, The five types of trump voters.
\newblock {\em\protect\JournalTitle{Democracy Fund Voter Study Group, June.
  Available (accessed 21 June 2017) at: https://www. voterstudygroup.
  org/reports/2016-elections/the-five-types-trump-voters}} (2017)
  \url{https://www.voterstudygroup.org/publication/the-five-types-trump-voters}.

\bibitem{delli2002internet}
MX Delli~Carpini, S Keeter, The internet and an informed citizenry.
\newblock {\em\protect\JournalTitle{Departmental Papers (ASC)}} p.~2 (2002).

\bibitem{lupia2000institutional}
A Lupia, MD McCubbins, The institutional foundations of political competence:
  How citizens learn what they need to know.
\newblock {\em\protect\JournalTitle{Elements of reason: Cognition, choice, and
  the bounds of rationality}} pp. 47--66 (2000).

\bibitem{yu2023partisanship}
X Yu, M Wojcieszak, A Casas, Partisanship on social media: in-party love among
  american politicians, greater engagement with out-party hate among ordinary
  users.
\newblock {\em\protect\JournalTitle{Political Behavior}} pp. 1--26 (2023).

\bibitem{mosleh2022field}
M Mosleh, G Pennycook, DG Rand, Field experiments on social media.
\newblock {\em\protect\JournalTitle{Current Directions in Psychological
  Science}} \textbf{31}, 69--75 (2022).

\bibitem{heissmatthes2019}
R Heiss, J Matthes, Does incidental exposure on social media equalize or
  reinforce participatory gaps? evidence from a panel study.
\newblock {\em\protect\JournalTitle{New Media \& Society}} \textbf{21},
  2463--2482 (2019).

\bibitem{Nanz_Matthes_2022}
A Nanz, J Matthes, Democratic consequences of incidental exposure to political
  information: A meta-analysis.
\newblock {\em\protect\JournalTitle{Journal of Communication}} \textbf{72},
  345–373 (2022).

\bibitem{Skovsgaard_Andersen_2020}
M Skovsgaard, K Andersen, Conceptualizing news avoidance: Towards a shared
  understanding of different causes and potential solutions.
\newblock {\em\protect\JournalTitle{Journalism Studies}} \textbf{21}, 459–476
  (2020).

\bibitem{Newman2019}
N Newman, Reuters institute digital news report 2019.
\newblock {\em\protect\JournalTitle{Reuters Institute for the study of
  Journalism}} p. 156 (2019).

\bibitem{Villi_Aharoni_Tenenboim-Weinblatt_Boczkowski_Hayashi_Mitchelstein_Tanaka_Kligler-Vilenchik_2022}
M Villi, et~al., Taking a break from news: A five-nation study of news
  avoidance in the digital era.
\newblock {\em\protect\JournalTitle{Digital Journalism}} \textbf{10}, 148–164
  (2022).

\bibitem{thorson2021algorithmic}
K Thorson, K Cotter, M Medeiros, C Pak, Algorithmic inference, political
  interest, and exposure to news and politics on facebook.
\newblock {\em\protect\JournalTitle{Information, Communication \& Society}}
  \textbf{24}, 183--200 (2021).

\bibitem{eady_how_2019}
G Eady, J Nagler, A Guess, J Zilinsky, JA Tucker, How {Many} {People} {Live} in
  {Political} {Bubbles} on {Social} {Media}? {Evidence} {From} {Linked}
  {Survey} and {Twitter} {Data}.
\newblock {\em\protect\JournalTitle{SAGE Open}} \textbf{9} (2019).

\bibitem{wojcieszak2020social}
M Wojcieszak, S Winter, X Yu, {Social Norms and Selectivity: Effects of Norms
  of Open-Mindedness on Content Selection and Affective Polarization}.
\newblock {\em\protect\JournalTitle{Mass Communication and Society}}
  \textbf{23}, 455--483 (2020).

\bibitem{Mukurjee2022}
S Mukerjee, K Jaidka, Y Lelkes, The political landscape of the u.s.
  twitterverse.
\newblock {\em\protect\JournalTitle{Political Communication}} \textbf{0}, 1--24
  (2022).

\bibitem{mcclain2021behaviors}
C McClain, R Widjaya, G Rivero, A Smith, The behaviors and attitudes of us
  adults on twitter.
\newblock {\em\protect\JournalTitle{Pew Research Center}} (2021).

\bibitem{tiktok21}
{WSJ Staff}, {Inside Tiktok's Highly Secretive Algorithm}
  (\url{https://www.wsj.com/video/series/inside-tiktoks-highly-secretive-algorithm/investigation-how-tiktok-algorithm-figures-out-your-deepest-desires/})
  (2021).

\bibitem{rossi2021closed}
WS Rossi, JW Polderman, P Frasca, The closed loop between opinion formation and
  personalized recommendations.
\newblock {\em\protect\JournalTitle{IEEE Transactions on Control of Network
  Systems}} \textbf{9}, 1092--1103 (2021).

\bibitem{nyhan2023like}
B Nyhan, et~al., Like-minded sources on facebook are prevalent but not
  polarizing.
\newblock {\em\protect\JournalTitle{Nature}} pp. 1--8 (2023).

\bibitem{Andersen2019softnews}
K Andersen, An entrance for the uninterested: Who watches soft news and how
  does it affect their political participation?
\newblock {\em\protect\JournalTitle{Mass Communication and Society}}
  \textbf{22}, 487--507 (2019).

\bibitem{Prior2003}
M Prior, Any good news in soft news? the impact of soft news preference on
  political knowledge.
\newblock {\em\protect\JournalTitle{Political Communication}} \textbf{20},
  149–171 (2003).

\bibitem{baum2006oprah}
MA Baum, AS Jamison, The oprah effect: How soft news helps inattentive citizens
  vote consistently.
\newblock {\em\protect\JournalTitle{The Journal of Politics}} \textbf{68},
  946--959 (2006).

\bibitem{baek2009don}
YM Baek, ME Wojcieszak, Don’t expect too much! learning from late-night
  comedy and knowledge item difficulty.
\newblock {\em\protect\JournalTitle{Communication Research}} \textbf{36},
  783--809 (2009).

\bibitem{moy2005communication}
P Moy, MA Xenos, VK Hess, Communication and citizenship: Mapping the political
  effects of infotainment.
\newblock {\em\protect\JournalTitle{Mass Communication \& Society}} \textbf{8},
  111--131 (2005).

\bibitem{Baum2010softnews}
M Baum, Soft news and political knowledge: Evidence of absence or absence of
  evidence?
\newblock {\em\protect\JournalTitle{Political Communication}} \textbf{20},
  173--190 (2010).

\bibitem{StierMangoldScharkowBreuer2022}
S Stier, F Mangold, M Scharkow, J Breuer, Post post-broadcast democracy? news
  exposure in the age of online intermediaries.
\newblock {\em\protect\JournalTitle{American Political Science Review}}
  \textbf{116}, 768–774 (2022).

\bibitem{wojcieszak2022avenues}
M Wojcieszak, E Menchen-Trevino, JF Goncalves, B Weeks, Avenues to news and
  diverse news exposure online: Comparing direct navigation, social media, news
  aggregators, search queries, and article hyperlinks.
\newblock {\em\protect\JournalTitle{The International Journal of
  Press/Politics}} \textbf{27}, 860--886 (2022).

\bibitem{Baum2003}
MA Baum, Soft news and political knowledge: Evidence of absence or absence of
  evidence?
\newblock {\em\protect\JournalTitle{Political Communication}} \textbf{20},
  173–190 (2003).

\bibitem{asr2021}
FT Asr, et~al., The gender gap tracker: Using natural language processing to
  measure gender bias in media.
\newblock {\em\protect\JournalTitle{PLoS ONE}} \textbf{1}, 173–190 (2021).

\bibitem{Sui2022}
M Sui, N Paul, C Hewitt, J Maki, K Searles, Is news for men?: Effects of
  women’s participation in news-making on audience perceptions and behaviors.
\newblock {\em\protect\JournalTitle{Journalism}} \textbf{0}, 14648849221125412
  (0).

\bibitem{Eckert2017}
E Stine, Fighting for recognition: Online abuse of women bloggers in germany,
  switzerland, the united kingdom, and the united states.
\newblock {\em\protect\JournalTitle{New Media and Society}} \textbf{20},
  1282--1302 (2018).

\bibitem{chen2020}
C GM, et~al., ‘you really have to have a thick skin’: A cross-cultural
  perspective on how online harassment influences female journalists.
\newblock {\em\protect\JournalTitle{Journalism}} \textbf{21}, 877--895 (2020).

\bibitem{Lewis2020}
SC Lewis, R Zamith, M Coddington, Online harassment and its implications for
  the journalist–audience relationship.
\newblock {\em\protect\JournalTitle{Digital Journalism}} \textbf{8}, 1047--1067
  (2020).

\bibitem{sobieraj2020credible}
S Sobieraj, {\em Credible threat: Attacks against women online and the future
  of democracy}.
\newblock (Oxford University Press), (2020).

\bibitem{usher2018twitter}
N Usher, J Holcomb, J Littman, Twitter makes it worse: Political journalists,
  gendered echo chambers, and the amplification of gender bias.
\newblock {\em\protect\JournalTitle{The international journal of
  press/politics}} \textbf{23}, 324--344 (2018).

\bibitem{naomi2020fedtos}
N Gilens, J Williams, {Federal Judge Rules It Is Not a Crime to Violate a
  Website’s Terms of Service. EFF}
  (\url{https://www.eff.org/deeplinks/2020/04/federal-judge-rules-it-not-crime-violate-websites-terms-service})
  (2020).

\bibitem{BotometerAPI}
M Sayyadiharikandeh, O Varol, KC Yang, A Flammini, F Menczer, Detection of
  novel social bots by ensembles of specialized classifiers in {\em Proceedings
  of the 29th ACM International Conference on Information \& Knowledge
  Management}, CIKM '20.
\newblock (Association for Computing Machinery, New York, NY, USA), p.
  2725–2732 (2020).

\bibitem{zhang2019dialogpt}
Y Zhang, et~al., Dialogpt: Large-scale generative pre-training for
  conversational response generation.
\newblock {\em\protect\JournalTitle{arXiv preprint arXiv:1911.00536}} (2019).

\bibitem{otero2019ad}
V Otero, Ad fontes media’s first multi-analyst content analysis ratings
  project (2019).

\bibitem{BERTPoliticalClassifier}
A Chhabra, Political classifier
  (\url{https://github.com/anshuman23/political_classifier}) (2023).

\bibitem{loureiro2022timelms}
D Loureiro, F Barbieri, L Neves, LE Anke, J Camacho-Collados, Timelms:
  Diachronic language models from twitter (2022).

\bibitem{barbieri2020tweeteval}
F Barbieri, J Camacho-Collados, L Neves, L Espinosa-Anke, Tweeteval: Unified
  benchmark and comparative evaluation for tweet classification (2020).

\bibitem{hainmueller_2012}
J Hainmueller, Entropy balancing for causal effects: A multivariate reweighting
  method to produce balanced samples in observational studies.
\newblock {\em\protect\JournalTitle{Political Analysis}} \textbf{20}, 25–46
  (2012).

\bibitem{chattonetal}
A Chatton, F Le~Borgne, L C., G-computation, propensity score-based methods,
  and targeted maximum likelihood estimator for causal inference with different
  covariates sets: a comparative simulation study.
\newblock {\em\protect\JournalTitle{Scientific Reports}} \textbf{10} (2020).

\bibitem{robins1986}
J Robins, A new approach to causal inference in mortality studies with a
  sustained exposure period—application to control of the healthy worker
  survivor effect.
\newblock {\em\protect\JournalTitle{Mathematical Modelling}} \textbf{7},
  1393--1512 (1986).

\bibitem{krupnikov2022other}
Y Krupnikov, JB Ryan, {\em The other divide}.
\newblock (Cambridge University Press), (2022).

\bibitem{wojcieszak2024non}
M Wojcieszak, et~al., Non-news websites expose people to more political content
  than news websites: Evidence from browsing data in three countries.
\newblock {\em\protect\JournalTitle{Political Communication}} \textbf{41},
  129--151 (2024).

\bibitem{wojcieszak2021no}
M Wojcieszak, et~al., No polarization from partisan news.
\newblock {\em\protect\JournalTitle{International Journal of Press and
  Politics}} (2021).

\bibitem{lupia1998democratic}
A Lupia, MD McCubbins, {\em The democratic dilemma: Can citizens learn what
  they need to know?}
\newblock (Cambridge University Press), (1998).

\bibitem{haroon}
X Yu, M Haroon, E Menchen-Trevino, M Wojcieszak, Nudging the recommendation
  algorithm increases news consumption and diversity on youtube (preprint
  \url{https://doi.org/10.21203/rs.3.rs-3349905/v1}) (2023).

\bibitem{twitterdemo}
{Digital 2023 October Global Statshot Report}
  (\url{https://datareportal.com/reports/digital-2023-october-global-statshot})
  (2023).

\bibitem{Pew2022}
S Bestvater, S Shah, G Rivero, A Smith, Politics on twitter: One-third of
  tweets from u.s. adults are political
  (\url{https://www.pewresearch.org/politics/2022/06/16/politics-on-twitter-one-third-of-tweets-from-u-s-adults-are-political/})
  (2022).

\bibitem{Pew2021}
A Mitchell, E Shearer, G Stocking, News on twitter: Consumed by most users and
  trusted by many
  (\url{https://www.pewresearch.org/journalism/2021/11/15/news-on-twitter-consumed-by-most-users-and-trusted-by-many/})
  (2021).

\bibitem{Pew2020}
P van Kessel, R Widjaya, S Shah, A Smith, A Hughes, Congress soars to new
  heights on social media
  (\url{https://www.pewresearch.org/internet/2020/07/16/congress-soars-to-new-heights-on-social-media/})
  (2020).

\bibitem{mcgregor2019social}
SC McGregor, Social media as public opinion: How journalists use social media
  to represent public opinion.
\newblock {\em\protect\JournalTitle{Journalism}} \textbf{20}, 1070--1086
  (2019).

\bibitem{barbera2019leads}
P Barber{\'a}, et~al., Who leads? who follows? measuring issue attention and
  agenda setting by legislators and the mass public using social media data.
\newblock {\em\protect\JournalTitle{American Political Science Review}}
  \textbf{113}, 883--901 (2019).

\bibitem{mosleh2021perverse}
M Mosleh, C Martel, D Eckles, D Rand, Perverse downstream consequences of
  debunking: Being corrected by another user for posting false political news
  increases subsequent sharing of low quality, partisan, and toxic content in a
  twitter field experiment in {\em proceedings of the 2021 CHI Conference on
  Human Factors in Computing Systems}.
\newblock pp. 1--13 (2021).

\bibitem{mosleh2021shared}
M Mosleh, C Martel, D Eckles, DG Rand, {Shared partisanship dramatically
  increases social tie formation in a Twitter field experiment}.
\newblock {\em\protect\JournalTitle{Proceedings of the National Academy of
  Sciences}} \textbf{118} (2021).

\bibitem{ventura2023whatsapp}
T Ventura, R Majumdar, J Nagler, JA Tucker, Whatsapp increases exposure to
  false rumors but has limited effects on beliefs and polarization: Evidence
  from a multimedia-constrained deactivation.
\newblock {\em\protect\JournalTitle{Available at SSRN 4457400}} (2023).

\bibitem{casas2022exposure}
A Casas, E Menchen-Trevino, M Wojcieszak, Exposure to extremely partisan news
  from the other political side shows scarce boomerang effects.
\newblock {\em\protect\JournalTitle{Political Behavior}} pp. 1--40 (2022).

\bibitem{guess2023social}
AM Guess, et~al., How do social media feed algorithms affect attitudes and
  behavior in an election campaign?
\newblock {\em\protect\JournalTitle{Science}} \textbf{381}, 398--404 (2023).

\bibitem{WellsThorson2018}
C Wells, K Thorson, Combining big data and survey techniques to model effects
  of political content flows in facebook.
\newblock {\em\protect\JournalTitle{Social Science Computer Review}}
  \textbf{35}, 33--52 (2017).

\bibitem{brown2021twitter}
ME Brown, PA Dustman, JJ Barthelemy, Twitter impact on a community trauma: An
  examination of who, what, and why it radiated.
\newblock {\em\protect\JournalTitle{Journal of Community Psychology}}
  \textbf{49}, 838--853 (2021).

\bibitem{Tweepy}
Tweepy documentation (\url{https://docs.tweepy.org/en/stable/api.html}) (2023).

\bibitem{Geostring}
D Freelon, {Geostring} (\url{https://github.com/dfreelon/geostring}) (2023).

\bibitem{SpacyLang}
M Honnibal, I Montani, S Van~Landeghem, A Boyd, {spaCy: Industrial-strength
  Natural Language Processing in Python} (2020).

\end{thebibliography}

\end{document}